\documentclass[aps,preprint]{revtex4}%
\usepackage{amsfonts}
\usepackage{amsmath}
\usepackage{amssymb}
\usepackage{graphicx}%
\begin{document}
\title{Surprising Symmetries in Relativistic Charge Dynamics }
\author{William E. Baylis}
\affiliation{Department of Physics, University of Windsor, Windsor, ON, Canada N9B 3P4}
\email{baylis@uwindsor.ca}
\date{}

\begin{abstract}
The eigenspinor approach uses the classical amplitude of the algebraic Lorentz
rotation connecting the lab and rest frames to study the relativistic motion
of particles. It suggests a simple covariant extension of the common
definition of the electric field: the electromagnetic field can be defined as
the proper spacetime rotation rate it induces in the particle frame times its
mass-to-charge ratio. When applied to the dynamics of a point charge in an
external electromagnetic field, the eigenspinor approach reveals surprising
symmetries, particularly the invariance of some field properties in the rest
frame of the accelerating charge. The symmetries facilitate the discovery of
analytic solutions of the charge motion and are simply explained in terms of
the geometry of spacetime. Symmetries of the uniformly accelerated charge and
electric dipole are also briefly discussed.

\end{abstract}
\maketitle

\section{Introduction}

The electric field is commonly defined as the acceleration per unit
charge-to-mass ratio of a small charge at rest. Lorentz transformations of the
charge and the covariant electromagnetic field then imply the Lorentz-force
equation. The eigenspinor (or rotor)
approach\cite{Hes74a,Bay99,DorLas03,Hes03b} to the motion of charged particles
in electromagnetic fields suggests a definition of the electromagnetic field
that is a simple extension of the textbook definition of the electric field.
The extension, which is required to make the equation of motion for the
eigenspinor linear, associates the electromagnetic field with the spacetime
rotation rate of any charge in the field. It implies not only the
Lorentz-force equation but also a spatial rotation in any magnetic field at
the cyclotron frequency.

A couple of rather surprising symmetries of the dynamics of charges in the
electromagnetic field result from the definition. The symmetries involve the
invariance of field in a classical \emph{rest frame }(commoving inertial
frame) of charges moving in (a) a uniform constant field, and (b) a pulsed
plane-wave field. This paper derives, examines, and illustrates these
symmetries, showing that they result directly from the definition of the field
and from the geometry of Minkowski spacetime. It also discusses symmetries in
the field of a uniformly accelerating charge and the interaction of a
uniformly accelerated physical dipole.

Symmetries simplify many physics problems and are often keys to finding
analytical solutions. This has in fact been demonstrated for charges in a
pulsed plane-wave field, where the symmetry, together with the powerful spinor
and projector tools inherent in the eigenspinor approach, has allowed analytic
solutions to be found\cite{Bay99a} for the relativistic motion of charged of
charged particles in propagating plane-wave pulses and in such pulses plus
constant longitudinal fields, such as occur, for example, in autoresonant
laser accelerators. It is important to understand the origin of the symmetries
so that their applications to other problems can be anticipated. This is
particularly true of relativistic symmetries that are often less obvious.

The eigenspinor is the amplitude of a Lorentz transformation relating the
instantaneously commoving particle frame with the lab.\cite{Bay99} It gives
directly both the proper velocity and relative orientation of the particle
frame and is related to the quantum wave function of the
particle.\cite{Bay92a} It arises as a natural rotor or transformation element
of the particle in the algebra of physical space (APS),\cite{Bay03} Clifford's
geometric algebra of vectors in three-dimensional Euclidean space.

In the following section, we review the spinorial form of Lorentz
transformations that arises naturally in treatments of classical relativistic
dynamics based on Clifford's geometric algebra. A summary of essential
features of APS is given in Appendix A. Section III describes the use of
rotors to describe the motion of charges in external fields, and Sections IV
and V present the dynamical symmetries for static and plane-wave fields,
respectively. A couple of further symmetries are briefly mentioned in Section
VI, followed by conclusions.

\section{Lorentz Rotors}

APS is the algebra of physical vectors and their products. It is isomorphic to
both complex quaternions and to the even subalgebra of the spacetime algebra,
and it is sometimes called the Pauli algebra because of its representation in
which the unit vectors $\mathbf{e}_{k}$ are replaced by the Pauli spin
matrices $\mathbf{\sigma}_{k}~.$ In fact, there are many possible matrix
representations, but only the algebra is important. The algebra is quite
simple, and we summarize its basic elements in the Appendix. More details can
be found elsewhere.\cite{Bay99,Bay03}

In the algebra, simple, physical (\textquotedblleft
restricted\textquotedblright) Lorentz transformations are rotations in
spacetime planes, and the Lorentz rotation of a spacetime vector $p$ such as
the momentum of a charge is given by the algebraic product%
\[
p\rightarrow LpL^{\dag},
\]
where $L=\exp\left(  \mathbf{W}/2\right)  $ is an amplitude of the rotation
called a Lorentz rotor, and $\mathbf{W}$ gives both the spacetime plane and
size of the rotation in that plane. The rotors are unimodular, that is unit
elements of the algebra: $L\bar{L}=1.$

\section{Eigenspinor}

The \emph{eigenspinor }$\Lambda$ of a particle is the Lorentz rotor $L$ that
transforms properties from the rest frame to the lab. By rest frame is meant
the commoving inertial frame. The inertial frame instantaneously at rest with
an accelerating particle is continuously changing. Given any known paravector
$p_{\text{rest}}$ in the commoving particle frame, it is transformed to the
lab by $p=\Lambda p_{\text{rest}}\Lambda^{\dag}.$ For example, the time axis
$\mathbf{e}_{0}$ in the rest frame becomes the dimensionless proper velocity
$u=\Lambda\mathbf{e}_{0}\Lambda^{\dag}$ and $p=mcu$ is the momentum of the particle.

The time development of the eigenspinor takes the form%
\begin{equation}
\dot{\Lambda}=\frac{1}{2}\mathbf{\Omega}\Lambda~, \label{eqmotion}%
\end{equation}
where $\mathbf{\Omega}=2\dot{\Lambda}\bar{\Lambda}$ is a spacetime plane
giving the \emph{spacetime rotation rate} of the particle and the dot
indicates a derivative with respect to the proper time. We note that
$\dot{\Lambda}$ and $\Lambda$ are orthogonal by virtue of the unimodularity of
$\Lambda.$ The proper-time derivative of the particle momentum is%
\[
\dot{p}=\frac{d}{d\tau}\left(  \Lambda mc\Lambda^{\dag}\right)  =\frac{1}%
{2}\left(  \mathbf{\Omega}p+p\mathbf{\Omega}^{\dag}\right)  \equiv\left\langle
\mathbf{\Omega}p\right\rangle _{\Re},
\]
where $\left\langle x\right\rangle _{\Re}=\left(  x+x^{\dag}\right)  /2$ is
the real (\emph{i.e.}, the hermitian) part of the element $x.$ The
Lorentz-force equation has exactly the same form:%
\begin{equation}
\dot{p}=\left\langle e\mathbf{F}u\right\rangle _{\Re}~, \label{LForce}%
\end{equation}
where $\mathbf{F}=\frac{1}{2}F^{\mu\nu}\left\langle \mathbf{e}_{\mu
}\mathbf{\bar{e}}_{\nu}\right\rangle _{V}=\mathbf{E}+ic\mathbf{B}$ is the
electromagnetic field. This follows from the eigenspinor equation of motion
(\ref{eqmotion}) for a spacetime rotation rate%
\begin{equation}
\mathbf{\Omega}=\frac{e}{mc}\mathbf{F.} \label{field}%
\end{equation}
and this rotation rate suggests an explicit, relativistically covariant
definition the electromagnetic field: $\mathbf{F}$ is the \emph{spacetime
rotation rate} of the frame of a classical point charge per unit $e/mc.$ For a
charge at rest, this definition reduces to the usual definition of the
electric field as the force per unit charge.

The identification (\ref{field}) is not the only relation between
$\mathbf{\Omega}$ and $\mathbf{F}$ that gives the Lorentz-force equation. The
Lorentz force is independent of the magnetic field in the rest frame, and
integration of the Lorentz-force equation gives the velocity and path of the
particle, but not the orientation of its frame. The choice (\ref{field})
satisfies the Lorentz force equation with a particular evolution of the
orientation of the particle frame. We call the frame to which it refers the
\emph{classical-particle frame}. It is the simplest choice for $\mathbf{\Omega
}$ consistent with the Lorentz-force equation (\ref{LForce}) and the only one
for which $\mathbf{\Omega}$ is independent of $\Lambda$ and for which the
equation of motion (\ref{eqmotion}) is therefore linear in $\Lambda.$ A more
general relation consistent with the Lorentz-force equation (\ref{LForce}) is
given in Appendix B, where it is shown that the choice (\ref{field})
corresponds to the nonspinning frame of a particle with a $g$-factor of 2.

\section{Surprising Symmetry 1: Charge in Uniform Field}

When $\mathbf{F}$ is any uniform, constant electromagnetic field, the solution
of (\ref{eqmotion},\ref{field}) is%
\[
\Lambda\left(  \tau\right)  =\exp\left(  \mathbf{\Omega}\tau/2\right)
\Lambda\left(  0\right)
\]
from which one generally gets a spacetime rotation (both a boost and a spatial
rotation) of the particle frame in the plane of $\mathbf{F.}$ The field seen
by the particle at proper time $\tau$ is
\begin{align*}
\mathbf{F}_{\text{rest}}\left(  \tau\right)   &  =\bar{\Lambda}\left(
\tau\right)  \mathbf{F}\Lambda\left(  \tau\right) \\
&  =\bar{\Lambda}\left(  0\right)  \exp\left(  -\mathbf{\Omega}\tau/2\right)
\mathbf{F}\exp\left(  \mathbf{\Omega}\tau/2\right)  \Lambda\left(  0\right) \\
&  =\mathbf{F}_{\text{rest}}\left(  0\right)
\end{align*}
since the plane $\mathbf{F}$ is invariant under rotations in the plane itself.
Thus the field seen by the accelerating charge is invariant!

This is hard to believe. Consider a familiar example: charge motion from rest
in crossed $\mathbf{E}$ and $\mathbf{B}$ fields when $c^{2}\mathbf{B}%
^{2}>\mathbf{E}^{2}$.%

\begin{figure}
[tbh]
\begin{center}
\includegraphics[
height=2.0245in,
width=3.3503in
]%
{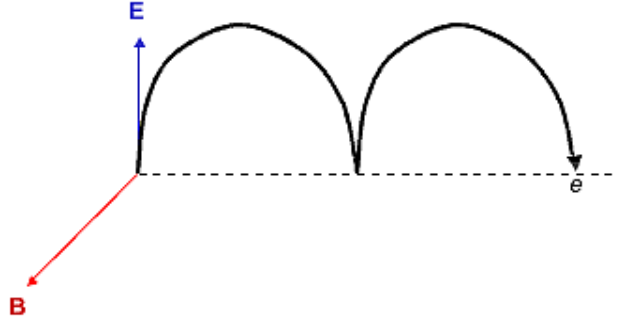}%
\caption{Cycloid motion of charge in crossed fields.}%
\label{cycloid}%
\end{center}
\end{figure}

At the top of the cycloid motion (see Fig. \ref{cycloid}), the charge is
moving at about twice the drift velocity (more precisely: $u=u_{\text{drift}%
}^{2}$ ) and the electric field in the unrotated charge frame has changed
sign. However, the particle frame has also rotated about the lab $\mathbf{B}$
direction by 180 degrees, so that the field it sees is unchanged. A similar
result can be shown for any point on the trajectory.

\section{Surprising Symmetry 2: Charge in Plane Wave}

Directed plane waves are null flags\cite{Pen84} of the form%
\begin{equation}
\mathbf{F=}\left(  1+\mathbf{\hat{k}}\right)  \mathbf{E}\left(  s\right)
,\label{nullflag}%
\end{equation}
where $\mathbf{\hat{k}}$ is the propagation direction, which is perpendicular
to the electric field $\mathbf{E,}$ assumed to be a known function of the
Lorentz scalar $s=\left\langle k\bar{x}\right\rangle _{S}=\omega
t-\mathbf{k\cdot x,}$ where $x$ is the spacetime position of the charge and
$k$ is a constant spacetime vector proportional to the spacetime wave vector.
The null-flag form (\ref{nullflag}) is imposed by Maxwell's equations for
source-free space, $\bar{\partial}\mathbf{F}=0,$ since for any nontrivial
field of the form $\mathbf{F}\left(  s\right)  $ they imply $\bar{k}%
\mathbf{F}=0,$ and this implies that both $k$ and $\mathbf{F}$ are
noninvertible and hence null: $k\bar{k}=0=\mathbf{F}^{2}.$ The spacetime wave
vector $k=\omega\left(  1+\mathbf{\hat{k}}\right)  /c$ is called the
\emph{flagpole} of the null flag.

\begin{figure}
[tbh]
\begin{center}
\includegraphics[
height=2.7605in,
width=3.685in
]%
{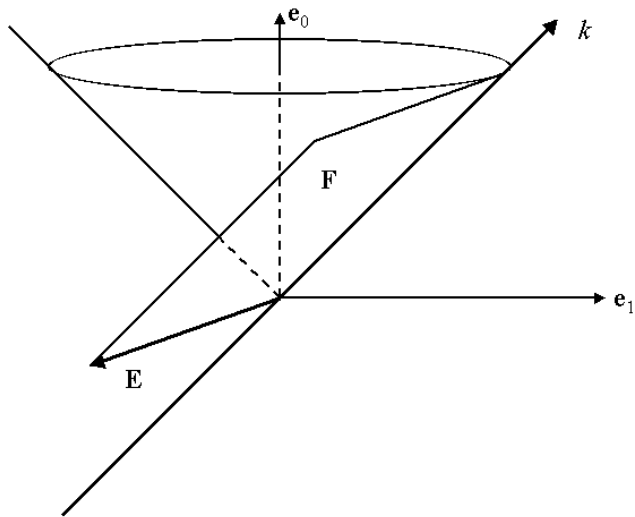}%
\caption{Null flag field. The spacetime wave vector $k$ is in the flag plane
but also orthogonal to it.}%
\label{flag}%
\end{center}
\end{figure}

The equation of motion (\ref{eqmotion},\ref{field}) can only be solved because
of a remarkable symmetry. In the particle rest frame it is%
\[
k_{\text{rest}}=\bar{\Lambda}k\bar{\Lambda}^{\dag}%
\]
which is a rotation of $k$ in the spacetime plane of $\mathbf{F.}$ But because
$k$ is null ( $k\bar{k}=0$ ), it is not only \emph{in} the null flag, it is
also \emph{orthogonal} to it, and therefore it is invariant under rotations in
$\mathbf{F}$ (see Fig. \ref{flag}). Thus, while the charge is being
accelerated by the plane wave, it continues to see a fixed wave paravector
$k_{\text{rest.}}$%

Consequently%
\[
\dot{s}=\left\langle k\bar{u}\right\rangle _{S}=\omega_{\text{rest}}
\]
is constant and the equation of motion reduces to%
\[
\dot{\Lambda}=\omega_{\text{rest}}\frac{d\Lambda}{ds}=\frac{e}{2mc}\left(
1+\mathbf{\hat{k}}\right)  \mathbf{E}\left(  s\right)  \Lambda\left(
0\right)
\]
which is trivially integrated. Note that since $\left(  1-\mathbf{\hat{k}%
}\right)  \dot{\Lambda}=0,$ we have
\begin{align*}
\left(  1+\mathbf{\hat{k}}\right)  \mathbf{E}\left(  s\right)  \Lambda\left(
\tau\right)   &  =\mathbf{E}\left(  s\right)  \left(  1-\mathbf{\hat{k}%
}\right)  \Lambda\left(  \tau\right)  =\mathbf{E}\left(  s\right)  \left(
1-\mathbf{\hat{k}}\right)  \Lambda\left(  0\right) \\
&  =\left(  1+\mathbf{\hat{k}}\right)  \mathbf{E}\left(  s\right)
\Lambda\left(  0\right)  .
\end{align*}

\section{Other Symmetries of the Electromagnetic Field}

\subsection{Boosts of Plane-Wave Fields}

Any two null flags (\ref{nullflag}) are related by a rotation and a dilation.
Since propagating plane waves have the form of a null flag, and any inertial
observer will see a propagating plane wave as a propagating plane wave, any
boost of a propagating plane wave must be equivalent to a spatial rotation and
dilation of that wave. This can be easily verified algebraically, where a less
obvious symmetry is also demonstrated: the boost applied to the spacetime wave
vector $k$ is also equivalent to a spatial rotation and dilation, and the
rotation angle and dilation factor are precisely the same as for the null
flag.\cite{Bay04c}

This symmetry allows one to derive results for waves obliquely incident on
plane conductors in terms of the normally incident case, and to express
wave-guide modes as boosted standing waves.

\subsection{Field of Uniformly Accelerated Charge}

Consider a point charge in hyperbolic motion tracing out the world line
$r\left(  \tau\right)  :$%
\begin{align}
r  &  =\exp\left(  c\tau\mathbf{e}_{3}\right)  =\gamma\left(  \mathbf{e}%
_{3}+\beta\right) \label{hyperbol}\\
u  &  \equiv c^{-1}\dot{r}=\mathbf{e}_{3}r=\gamma\left(  1+\beta\mathbf{e}%
_{3}\right)  ~.\nonumber
\end{align}
Here, $r^{0}\equiv ct=\gamma\beta=\sinh c\tau$ is the local coordinate time,
$\beta$ is the speed of the charge in units of $c,$ and $\gamma=\left(
1-\beta^{2}\right)  ^{-1/2}.$ The unit of length is $l_{0}=c^{2}/\left\vert
\mathbf{a}_{r}\right\vert =1,$ where $\mathbf{a}_{r}=c\dot{u}\bar{u}%
=c^{2}\mathbf{e}_{3}$ is the acceleration in the rest frame. We note $u\bar
{u}=1=-r\bar{r}$ and $r\bar{u}=\mathbf{e}_{3}.$ The field position is
$x=x^{\mu}\mathbf{e}_{\mu}=x_{0}+\mathbf{x,}$ and the relative position
$R=x-r$ is lightlike:
\[
R\bar{R}=R_{0}^{2}-\mathbf{R}^{2}=x\bar{x}-1-2\left\langle x\bar
{r}\right\rangle _{S}=0\,.
\]
This is the retarded condition that gives the retarded proper time $\tau$ in
terms of $x.$The Li\'{e}nard-Wiechert field\cite{Bay99}
\[
\mathbf{F}=\frac{Kce}{\left\langle x\bar{u}\right\rangle _{S}^{3}}\left(
\left\langle R\bar{u}\right\rangle _{V}+\frac{1}{2c}R\overline{\dot{u}}%
u\bar{R}\right)
\]
is the sum of the boosted Coulomb field and the acceleration field. Neither
field by itself satisfies Maxwell's equations; only the sum does. For the
hyperbolic motion (\ref{hyperbol}), the total field reduces to%
\[
\mathbf{F}=-\frac{Kce}{2\left\langle x\bar{u}\right\rangle _{S}^{3}}\left(
\mathbf{e}_{3}+x\mathbf{e}_{3}\bar{x}\right)  \,.
\]
At the instant $t=0,$ $x=\mathbf{x}$ and $\mathbf{F}$ is purely real: the
magnetic field vanishes throughout space.

The boosted Coulomb field and the acceleration field separately have magnetic
parts, but their sum cancels everywhere at $t=0$, as it must by the
equivalence principle. It is surprising that part of the essentially $R^{-1}$
radiation field can be canceled by the Coulomb term. The usual $R^{-2}$
dependence of the Coulomb term has a $R^{-1}$ behavior because the proper
velocity $u$ at the retarded time grows linearly in $R,$ and this makes the
cancellation possible. The electric field lines are curved away from the
direction of acceleration. In the equivalent case of the uniform gravitational
field, one would ascribe the curvature of the field lines to being a result of
the gravitational field. The interesting implications for the interpretation
of of the radiative field will be explored elsewhere.

The dynamical symmetry in this case arises when we consider the interaction of
two opposite charges that form a dipole. Both charges have the same hyperbolic
motion (\ref{hyperbol}) but are held displaced a small distance from each
other along a direction perpendicular to $\mathbf{e}_{3}$. At $t=0,$ both
charges are instantaneously at rest and each interacts with the purely
electric field of the other. However, because of the curvature of the electric
field lines, there is a net force in the direction of the acceleration. This
force is readily evaluated and corresponds to a reduction in the gravitational
force on the dipole that it would experience in a gravitational field. One
also expects a reduction in the mass of the dipole arising from the attractive
electromagnetic interaction between the charges. It is easily confirmed that
at small separations the resultant reduction in gravitational force equals to
the upward lift arising from the curved electric-field lines.

\section{Conclusion}

Much beautiful symmetry in electrodynamics can best be appreciated in a
relativistically covariant formulation of electrodynamics such as APS that
emphasizes the geometry of spacetime while maintaining clear relationships to
the space and time components seen by any observer. The extension of the
definition of the electric field to the covariant electromagnetic field
$\mathbf{F}$ in terms of the spacetime rotation rate of the classical
charged-particle frame leads to new symmetries that are powerful tools for
solving some problems in relativistic dynamics.

\subsection*{Acknowledgment}

The author thanks the Natural Sciences and Engineering Research Council of
Canada for support of this research.

\section*{Appendix A: Summary of APS}

The structure of APS is entirely determined by the axiom that the square of
any vector is its length squared: $\mathbf{v}^{2}\equiv\mathbf{vv}%
=\mathbf{v\cdot v,}$ together with the usual associative and distributive laws
for the sums and products of square matrices. For example, it follows directly
from the axiom that aligned vectors, which are proportional to each other,
commute, and that every nonzero vector has a inverse: $\mathbf{v}%
^{-1}=\mathbf{v}/\mathbf{v}^{2}.$ In particular, unit vectors such as
$\mathbf{e}_{k},~k=1,2,3,$ are their own inverses, and an explicit operator
that transforms the vector $\mathbf{v}$ into $\mathbf{w}$ can be written
$\mathbf{wv}^{-1}$. By replacing $\mathbf{v}$ in the axiom by the sum of
perpendicular vectors, one sees that perpendicular vectors anticommute.
Indeed, $\mathbf{e}_{2}\mathbf{e}_{1}=-\mathbf{e}_{1}\mathbf{e}_{2}$ is called
a bivector, and when operating from the left on any vector $\mathbf{v}%
=v_{x}\mathbf{e}_{1}+v_{y}\mathbf{e}_{2}$ in the $\mathbf{e}_{1}\mathbf{e}%
_{2}$ plane, it gives another vector, related to the original by a $\pi/2$
rotation in the plane: $\mathbf{e}_{2}\mathbf{e}_{1}\mathbf{v}=v_{x}%
\mathbf{e}_{2}-v_{y}\mathbf{e}_{1}$. To rotate $\mathbf{v}$ by the angle
$\phi$ in the plane, we can multiply by $\exp\left(  \mathbf{e}_{2}%
\mathbf{e}_{1}\phi\right)  =\cos\phi+\mathbf{e}_{2}\mathbf{e}_{1}\sin\phi,$
where the Euler relation follows by power-series expansions when one notes
that $\left(  \mathbf{e}_{2}\mathbf{e}_{1}\right)  ^{2}=-1.$ The bivector
$\mathbf{e}_{2}\mathbf{e}_{1}$ thus generates rotations in the $\mathbf{e}%
_{2}\mathbf{e}_{1} $ plane. A rotation of a general vector $\mathbf{u}%
=u_{x}\mathbf{e}_{1}+u_{y}\mathbf{e}_{2}+u_{z}\mathbf{e}_{3}$ by $\phi$ in the
$\mathbf{e}_{2}\mathbf{e}_{1}$ plane can be expressed by what is called a spin
transformation\cite{Loun97}%
\[
\mathbf{u}\rightarrow R\mathbf{u}R^{\dag},
\]
where $R=\exp\left(  \mathbf{e}_{2}\mathbf{e}_{1}\phi/2\right)  $ is a rotor
and $R^{\dag}=\exp\left[  \left(  \mathbf{e}_{2}\mathbf{e}_{1}\right)  ^{\dag
}\phi/2\right]  =\exp\left(  \mathbf{e}_{1}\mathbf{e}_{2}\phi/2\right)  $ is
its \emph{reversion,} obtained by reversing the order of vector factors. The
notation reflects the fact that in any matrix rotation in which the basis
vectors are hermitian, reversion corresponds to hermitian conjugation. Note
that in this sense, $R$ is also unitary: $R^{\dag}=R^{-1},$ and it follows
that the bivector $\mathbf{e}_{2}\mathbf{e}_{1}$ is itself invariant under
rotations in the $\mathbf{e}_{2}\mathbf{e}_{1}$ plane.

A general element of APS can contain scalar, vector, bivector, and trivector
parts. The unit trivector $\mathbf{e}_{1}\mathbf{e}_{2}\mathbf{e}_{3}$ is
invariant under any rotations, commutes with vectors and hence all elements of
APS, and squares to $-1.$ It is called the pseudoscalar of the algebra and can
be identified with the unit imaginary $i.$ The linear space of APS is thus
eight-dimensional space over the reals and contains several subspaces,
including the original vector space, spanned by $\left\{  \mathbf{e}%
_{1},\mathbf{e}_{2},\mathbf{e}_{3}\right\}  ,$ the bivector space, spanned by
$\left\{  \mathbf{e}_{2}\mathbf{e}_{3}\mathbf{,e}_{3}\mathbf{e}_{1}%
\mathbf{,e}_{1}\mathbf{e}_{2}\right\}  ,$ the complex field, spanned by
$\left\{  1,\mathbf{e}_{1}\mathbf{e}_{2}\mathbf{e}_{3}=i\right\}  ,$ and
direct sums of subspaces such as paravector space, spanned by $\left\{
1,\mathbf{e}_{1},\mathbf{e}_{2},\mathbf{e}_{3}\right\}  .$ Paravectors are
sums of scalars plus vectors. The linear space of APS can be viewed as
paravector space over the complex field.

The Euclidean metric of physical space induces a Minkowski spacetime metric on
paravector space. This is seen by noting that the square of a paravector
$p=p^{0}+\mathbf{p}$ is generally not a scalar, but that $p\bar{p}=\left(
p^{0}\right)  ^{2}-\mathbf{p}^{2}$ always is, where $\bar{p}\equiv
p^{0}-\mathbf{p}$ is the Clifford conjugate of $p$. It is convenient to denote
$\mathbf{e}_{0}=1$ so that paravectors can be written with the Einstein
summation convention as $p=p^{\mu}\mathbf{e}_{\mu}.$ The scalar product of
paravectors $p$ and $q$ is then the scalar-like part of the product $p\bar
{q}:$%
\[
\left\langle p\bar{q}\right\rangle _{S}\equiv\frac{1}{2}\left(  p\bar{q}%
+q\bar{p}\right)  =p^{\mu}q^{\nu}\eta_{\mu\nu}
\]
where the metric tensor $\eta_{\mu\nu}=\left\langle \mathbf{e}_{\mu
}\mathbf{\bar{e}}_{\nu}\right\rangle _{S}$ is exactly that of Minkowski
spacetime. It is therefore natural to use paravectors as spacetime vectors,
where the scalar part of the paravector represents the time component of the
spacetime vector. If $\left\langle p\bar{q}\right\rangle _{S}=0,$ the
paravectors $p$ and $q$ are orthogonal to each other. The inverse of a
paravector $p$ is $p^{-1}=\bar{p}/p\bar{p},$ but this exists only if $p$ is
not null: $p\bar{p}\neq0.$ An explicit algebraic operator that transforms $p $
into $q$ is $qp^{-1}=q\bar{p}/p\bar{p}.$

Any two non-collinear paravectors $p,q$ determine a plane in spacetime
represented by the biparavector $\left\langle p\bar{q}\right\rangle _{V}%
\equiv\frac{1}{2}\left(  p\bar{q}-q\bar{p}\right)  .$ Biparavector space is
six dimensional and is spanned by $\left\{  \left\langle \mathbf{e}_{\mu
}\mathbf{\bar{e}}_{\nu}\right\rangle _{V}\right\}  _{0\leq\mu<\nu\leq3}.$ It
is a direct sum of the vector and bivector spaces of APS. Biparavectors
generate rotations in spacetime, and these are the physical (restricted)
Lorentz transformations, which we may also call Lorentz rotations. Such
rotations preserve the scalar product of paravectors and can be generally
written in the same form as a spatial rotation:%
\[
p\rightarrow LpL^{\dag},
\]
where $L=\pm\exp\left(  \mathbf{W}/2\right)  $ is a Lorentz rotor and
$\mathbf{W}=\frac{1}{2}W^{\mu\nu}\left\langle \mathbf{e}_{\mu}\mathbf{\bar{e}%
}_{\nu}\right\rangle _{V}$ is a biparavector. If $\mathbf{W}$ contains only
bivector parts, $L=\bar{L}^{\dag}$ is unitary and gives a spatial rotation; if
$\mathbf{W}$ contains only vector parts, $L=L^{\dag}$ and gives a boost. In
all cases, $L$ is unimodular: $L\bar{L}=1.$

Because of the unimodularity, the Lorentz rotation of a biparavector takes the
form%
\[
p\bar{q}\rightarrow LpL^{\dag}\overline{\left(  LqL^{\dag}\right)  }=Lp\bar
{q}\bar{L}
\]
and in particular, the biparavector for the spacetime plane of a Lorentz
rotation is invariant under that rotation.

\section*{Appendix B: General Rotation Rate for a Spinning Particle}

As discussed above, the Lorentz-force equation determines the path of a
charge, starting with a given position and velocity, in an electromagnetic
field $\mathbf{F,}$ but it does not give the orientation of its frame. The
eigenspinor $\Lambda$ gives both the path and the orientation. Mathematically,
the acceleration $\dot{u}$ depends only on the real part of $\mathbf{\Omega}$
in the particle rest frame, whereas $\Lambda$ depends on both the real and
imaginary parts. However, if the charge also possesses a spin with an
associated magnetic moment, that spin will precess in a magnetic field, and
this precession constrains the evolution of the orientation. The most general
relation consistent with the Lorentz-force equation (\ref{LForce}) can be
expressed%
\begin{equation}
\mathbf{\Omega}=\frac{e}{mc}\left[  \mathbf{F}+\frac{g-2}{4}\left(
\mathbf{F}-u\mathbf{F}^{\dag}\bar{u}\right)  \right]  +\omega_{0}\mathbf{S~,}
\label{Omega}%
\end{equation}
where $g$ is the $g$-factor for the spin and $\omega_{0}$ is the spin rate in
the spin plane given by the unit bivector $\mathbf{S.}$ This can be
shown\cite{Bay04b} to give the well-known BMT equation\cite{BMT59} for the
motion of a classical point charge with spin. The last term on the RHS
represents the rotation rate associated with the spin, while the second term
on the RHS\ contains a $u$ dependence that makes the eigenspinor equation of
motion (\ref{eqmotion}) nonlinear. The simpler result (\ref{field}) is the
spacetime rotation rate of what we may call the \emph{classical particle
frame:} a non-spinning frame tied to the point charge. Because $\mathbf{\Omega
}$ (\ref{field}) does not depend on the velocity of the charge, it gives a
linear equation of motion (\ref{eqmotion}). By comparison to the general
expression (\ref{Omega}), the spacetime rotation rate classical particle frame
is that of a particle with $\omega_{0}=0$ and $g=2.$

\bibliographystyle{apsrev}
\bibliography{bay,geomalg,phystxts}

\end{document}